\documentclass{emulateapj}
\slugcomment{Published in PASP, 119, 1449 (Dec. 2007)}
\shorttitle{Reddening in SDSS Photometric System}
\shortauthors{KIM \& LEE}

\def\spose#1{\hbox to 0pt{#1\hss}}
\newcommand\lsim{\mathrel{\spose{\lower 3.0pt\hbox{$\mathchar"218$}}
     \raise 2.0pt\hbox{$\mathchar"13C$}}}
\newcommand\gsim{\mathrel{\spose{\lower 3.0pt\hbox{$\mathchar"218$}}
     \raise 2.0pt\hbox{$\mathchar"13E$}}}
\newcommand\msun{{\rm \,M_\odot}}

\begin{document}
\title{Reddening Behaviors of Galaxies in the SDSS Photometric System}
\author{Sungsoo S. Kim\altaffilmark{1} and Myung Gyoon Lee\altaffilmark{2}}
\altaffiltext{1}{Dept. of Astronomy \& Space Science, Kyung Hee University,
Yongin-shi, Kyungki-do 449-701, Korea; sungsoo.kim@khu.ac.kr}
\altaffiltext{2}{Astronomy Program, Dept. of Physics \& Astronomy,
Seoul National University, Seoul 151-742, Korea; mglee@astrog.snu.ac.kr}

\begin{abstract}
We analyze the behaviors of reddening vectors in the SDSS photometric system
for galaxies of different morphologies, ages, and redshifts.  As seen
in other photometric systems, the dependence of reddening on the spectral
energy distribution (SED) and the nonlinearity of reddening are
likewise non-negligible for the SDSS system if extinction is significant
($\gsim 1$~mag).  These behaviors are most significant for the $g$ filter,
which has the largest bandwidth-to-central wavelength ratio among SDSS
filters.  The SDSS colors involving adjacent filters show greater
SED-dependence and nonlinearity.  A procedure for calculating the
correct amount of extinction from an observed color excess is provided.
The relative extinctions between (i.e., the extinction law for) SDSS filters
given by Schlegel et al., which were calculated with an older version of
filter response functions, would underestimate the amount of extinction
in most cases by $\sim 5$ to 10~\% (maximum $\sim 20$~\%).  We recommend
$A/A_{5500\AA}$ values of 1.574, 1.191, 0.876, 0.671, \& 0.486 for the
$u$, $g$, $r$, $i$, \& $z$ filters, respectively, as a representative
extinction law for the SDSS galaxies with a small extinction (i.e., for
cases where the nonlinearity and SED-dependence of the reddening is not
important).  The dependence of reddening on redshift at low extinction
is the largest for colors involving the $g$ filter as well, which is due
to the Balmer break.
\end{abstract}

\section{INTRODUCTION}
\label{sec:introduction}

The Sloan Digital Sky Survey (SDSS; Stoughton et al. 2002) completed its
first phase of operations in June 2005, and has entered a second phase,
SDSS-II, which will continue until June 2008.  One of the three surveys being
carried out by SDSS-II is the Sloan Extension for Galactic Understanding and
Exploration (SEGUE), which is currently mapping the structure of the Milky Way.
SEGUE will produce 3,500 square degrees of new images, some on a regular
grid extending through the Galactic plane and some sampling structures
like the Sagittarius dwarf tidal stream.  With the information from higher
to lower Galactic latitudes sampled on a uniform longitudinal grid,
SEGUE will significantly improve our understanding of the thick and thin
disks of the Milky Way, as well as of the external galaxies in the zone of
avoidance.  

The analyses of photometric data from the lower Galactic latitudes, however,
will require careful step of removing the effects of the Galactic interstellar
extinction.
One of the most common ways of estimating the amount of interstellar
extinction from photometric data involves a color-magnitude (CM) or
color-color (CC) diagram: one measures the amount of color excess and converts
it to an extinction value by using a certain extinction law.
This procedure seems simple, but care must
be taken when the extinction is significant, because the amount of
extinction in magnitudes of a certain photometric band depends on the detailed
spectral energy distribution (SED) of the object and is moreover not simply
proportional to the amount of interstellar dust.  The former effect makes the
slope of the reddening vector in the CM or CC diagrams dependent on the type
of the object, while the latter makes the reddening vector nonlinear.  These
effects arise because the transmission functions of the photometric filters
have non-negligible bandwidths (i.e. they are not Dirac delta functions),
and they hence become more important for wide-band filters.

The first thorough study on the SED dependence and nonlinearity of
interstellar extinction was performed by Grebel \& Roberts (1995; see
the references therein for earlier works on these issues).  Using the
Kurucz models of synthetic stellar spectra, they derived color-dependent
interstellar extinction relations for Johnson-Cousins $UBVRI$ and
Washington $CMT_1T_2$ photometric systems.  They found that the reddenings
of main-sequence stars and evolved stars are quite different and also can
be nonlinear as functions of color.

While Grebel \& Roberts (1995) studied the reddening behaviors of individual
stars in the optical bands, Kim et al. (2005) and Kim, Figer,
\& Lee (2006) investigated the collective behaviors of the reddening for
various isochrones in the near-infrared CM diagrams.  They found that
the reddened isochrones of different ages and metallicities behave as if
they follow different extinction laws, and that the reddening vectors of
some filter pairs in the CM diagrams (particularly those in the {\it Hubble
Space Telescope} NICMOS) are considerably nonlinear.  They provided a so-called
``effective extinction slope'' for each filter pair and isochrone model as well
as the coefficients of the third-order polynomials that best fit the reddening
vectors in the NICMOS CM diagrams for various isochrones.

There have been some theoretical studies on the extinction properties of
the SDSS photometric system.  Schlegel, Finkbeiner, \& Davis (1998)
and Stoughton et al. (2002), among others, have calculated the extinction
ratios $A_i/A_V$ of five SDSS bands ($u$, $g$, $r$, $i$, and $z$) relative
to the $V$ band for a normal elliptical galaxy SED.  Fiorucci \&
Munari (2003) provided the minimum and maximum values of $A_i/A_V$
for several different stars.
The dependence of extinction on stellar parameters and the nonlinearity of
extinction for SDSS bands were studied by Girardi et al. (2004; GGOC hereafter).
Although their main goal was to provide the theoretical isochrones in the SDSS
system, they also gave $A_i/A_V$ values for stars with a few different
values of effective temperature $T_{eff}$, surface gravity $\log g$, and
metallicity [M/H] as well.  They calculated $A_i/A_V$ values for different
$A_V$ values of up to 3~mag as well, but only for one set of stellar
parameters.  The study by GGOC clearly shows that SED dependence and
nonlinearity of extinction are non-negligible for SDSS bands as well.

In the present paper, we extend the study by GGOC and analyze the
behavior of reddening in the SDSS photometric system for SEDs of several
representative galaxies.  We show that the nonlinearity and the SED-dependence
of the reddening is not negligible for the SDSS filters.  Our results are
to be used for correctly estimating the amount of extinction.
Due to the SED-dependence of the reddening, one needs
to know the approximate morphology, age, and redshift of the galaxy before
estimating the extinction.  For cases where the spectrum or visual
classification of morphology is not available, we discuss how one could
estimate such information from the photometry.

This paper is composed as follows.  We describe our models in
\S~\ref{sec:model}, present and discuss our calculations in
\S~\ref{sec:reddening}, and summarize our findings in \S~\ref{sec:summary}.

\section{MODELS}
\label{sec:model}

\subsection{Magnitudes}

Following Fukugita et al. (1996) and GGOC, we adopt an AB magnitude system
(Oke \& Gunn 1983), which has a flat reference spectrum of flux $f_\nu^0
=f_\lambda^0 \lambda^2 /c =3.631 \times 10^{-20}$ erg s$^{-1}$ cm$^{-2}$
Hz$^{-1}$ and a reference
magnitude $m^0_{S_\lambda} = 0$ for all filters.  The apparent magnitude
$m_{S_\lambda}$, as measured using a filter with transmission function
$S_\lambda$, is given by
\begin{equation}
\label{mag}
        m_{S_\lambda} = -2.5 \log \left (
                        \frac{\int \lambda f_\lambda S_\lambda d \lambda}
                             {\int \lambda f_\nu^0 S_\lambda d \lambda}
                        \right ) + m_{S_\lambda}^0,
\end{equation} 
where the flux of a star measured at the top of the Earth atmosphere,
$f_\lambda$, is related to the flux at the stellar photosphere, $F_\lambda$, by
\begin{equation}
        f_\lambda = 10^{-0.4 A_\lambda} (R/d)^2 F_\lambda.
\end{equation}
Here $A_\lambda$ is the extinction in magnitudes, $R$ the stellar radius,
and $d$ the distance to the star.

We use the SDSS filter response functions for airmass $X=1.3$ available from
the SDSS Data Release 5 Web
pages.\footnote{http://www.sdss.org/dr5/instruments/imager/filters/index.html.}
The central wavelength
\begin{equation}
\label{lambda_c}
	\lambda_{c} = \exp \frac{\int d(\ln \nu) \, S_\nu \ln \lambda}
	                       {\int d(\ln \nu) \, S_\nu}
\end{equation}
(defined by Schneider, Gunn, \& Hoessel 1983 as the effective wavelength)
and the bandwidth
\begin{equation}
\label{bandwidth}
	\Delta \lambda = \frac{\int d\lambda \, S_\lambda}
	                      {\rm{max} \, \, S_\nu}
\end{equation}
of SDSS filters are presented in Table \ref{table:lambda}.
When convolving the SED of isochrones and galaxies with the transmission
curve, we adopt the photon integration (eq. \ref{mag}) instead of the
energy integration (eq. \ref{mag} without $\lambda$ in both integrands)
as data from the charge-coupled devices would be better represented by
the former.

\begin{deluxetable}{lccccc}
\tablecolumns{6}
\tablewidth{0pt}
\tablecaption{
\label{table:lambda}SDSS Filter Properties}
\tablehead{
\colhead{Filter} &
\colhead{$u$} &
\colhead{$g$} &
\colhead{$r$} &
\colhead{$i$} &
\colhead{$z$}
}
\startdata
$\lambda_{c}$ (\AA)            &  3588 &  4797 &  6232 &  7549 &  9039  \\
$\Delta \lambda$ (\AA)         &   558 &  1158 &  1111 &  1041 &  1125  \\
$\Delta \lambda / \lambda_{c}$ & 0.156 & 0.241 & 0.178 & 0.138 & 0.124  \\
\enddata
\tablecomments{$\lambda_{c}$ and $\Delta \lambda$ are the central wavelength
(eq. \ref{lambda_c}) and the bandwidth (eq. \ref{bandwidth}) of the filter,
respectively.}
\end{deluxetable}

\subsection{Galaxies}

\begin{figure*}
\epsscale{0.8}
\plotone{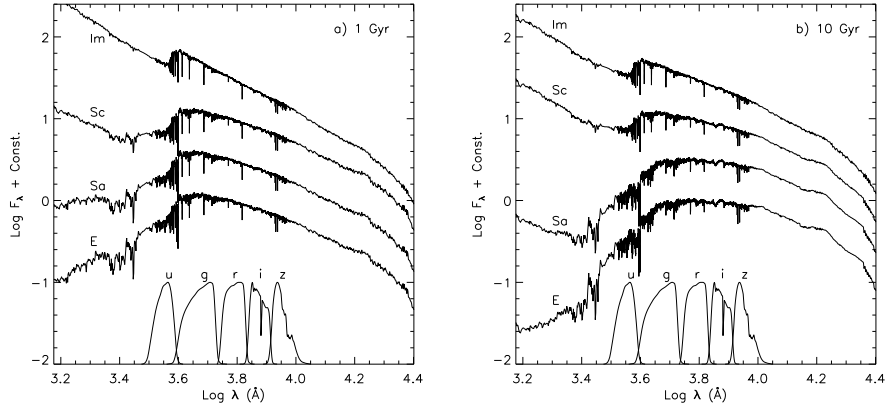}
\caption
{\label{fig:galspec}Spectra of our $z=0$ galaxy models for four different
morphologies at 1~Gyr ($a$) and 10~Gyr ($b$).}
\end{figure*}

We adopt the stellar population synthesis model by Bruzual \& Charlot (2003)
for the SEDs of simple stellar populations (SSPs; or instantaneous-burst
models) computed using the initial mass function (IMF) given by Chabrier (2003)
with lower and upper mass cutoffs of 0.1 and $100 \msun$, respectively.
Then these SSPs are combined to form the SEDs of four different galaxy
morphologies, E, Sa, Sb, \& Im, following the recipes by Buzzoni (2005).
Buzzoni (2005) implemented three [Fe/H] values to build up his galaxy models
but those values do not coincide with those used in the SSP models given
by Bruzual \& Charlot (2003).  So, we interpolate the SSP
SEDs for the [Fe/H] values used by the Buzzoni models.
Figure \ref{fig:galspec} shows the SEDs for 3 and 10 Gyr template galaxies of
four different morphologies constructed in this way.\footnote{An electronic
version of our galaxy SEDs is available from the authors upon request.}

\subsection{The Extinction Law}

We adopt the extinction law by Cardelli, Clayton, \& Mathis (1989) for a
typical Galactic total-to-selective ratio of $R_V \equiv A_V/(A_B-A_V) = 3.1$.
They provide fitting formulae to calculate the amount of interstellar
extinction at wavelength $\lambda$ relative to the one at the $V$ filter
($A_\lambda/A_V$).  But as $A_V$ also exhibits the SED dependence
and nonlinearity, we regard their extinction law as the extinction relative
to the one at $\lambda = 5500$~{\AA} ($A_\lambda/A_{5500}$), which is the pivot
wavelength of their extinction formulae.  We find that our $A_\lambda/A_{5500}$
values for select synthetic stellar spectra agree with those of
GGOC within 1~\% (it appears that the $A_\lambda/A_V$ values of GGOC is
defined in the same way as our $A_\lambda/A_{5500}$).

\section{REDDENING BEHAVIORS}
\label{sec:reddening}

Now we analyze the reddening behaviors of various galaxy models
for $A_\lambda$ values of up to 5 mag in each SDSS band.
We calculate how the magnitudes of our galaxy models change as we increase
$A_{5500}$ and obtain the actual extinction values, $A^{act}$, for each SDSS
filter.  Then the differences and ratios between these $A^{act}$ values for
each galaxy model determine the reddening curves in the CM and CC diagrams
(we herein use the term ``reddening curve'' instead of
``reddening vector'' to emphasize the fact that the actual reddening is
not linear in the CM and CC diagrams; we use the term reddening vector only
for the ``reference reddening vector'' described below, which we mean to be
linear).

\subsection{Reddening Estimation from Color Excess}

For an object whose intrinsic color can be estimated,
the amount of extinction is estimated by transforming an observed
color excess to the amount of extinction following a certain extinction
law.  It is customary to assume that the reddening
is linear and not dependent on the type of the object,
and the transformation is usually carried out with
an assumption that a single, linear reddening vector for a given
filter set can be applied to various objects with different SEDs.
To quantify the error involved in the extinction estimation using
these two assumptions, we define a ``reference extinction law''
as the amount of extinction relative to 5500~{\AA} ($A_i/A_{5500}$)
that an imaginary filter having a Dirac delta transmission function at
$\lambda_c$ of an SDSS filter $i$ would experience for the SED of
a 10~Gyr elliptical galaxy at redshift $z$ of 0 when $A_{5500}=0.1$.
Table \ref{table:Ai} gives $A_i/A_{5500}$ values for the SDSS filters
and the reddening vectors calculated from this extinction law will be
called ``reference reddening vectors.''

\begin{deluxetable}{lllllll}
\tablecolumns{6}
\tablewidth{0pt}
\tablecaption{
\label{table:Ai}Relative Extinction for SDSS Filters}
\tablehead{
\colhead{} &
\colhead{} &
\colhead{$u$} &
\colhead{$g$} &
\colhead{$r$} &
\colhead{$i$} &
\colhead{$z$}
}
\startdata
Our Reference\tablenotemark{a}
    & & 1.574 & 1.191 & 0.876 & 0.671 & 0.486 \\
Schlegel et al.\tablenotemark{b}
    & & 1.579 & 1.161 & 0.843 & 0.639 & 0.453 \\
\enddata
\tablenotetext{a}{$A_i/A_{5500}$ values for the SED of a 10~Gyr,
elliptical galaxy at redshift $z=0$ when $A_{5500}=0.1$.}
\tablenotetext{b}{$A_i/A_V$ values calculated by Schlegel et al.
(1998) for a normal elliptical galaxy SED.}
\end{deluxetable}

The amount of extinction estimated from a calculated color
excess and the reference reddening vector will be labeled by $A^{est}$
(this is our ``simulated'' amount of extinction that one would estimate
from observed color excess in practice when not considering the SED-dependence
and nonlinearity of the reddening).  In our analyses below, we will
describe the behaviors of reddening in terms of the values $(A^{est}-A^{act})$.

\begin{figure*}
\epsscale{0.8}
\plotone{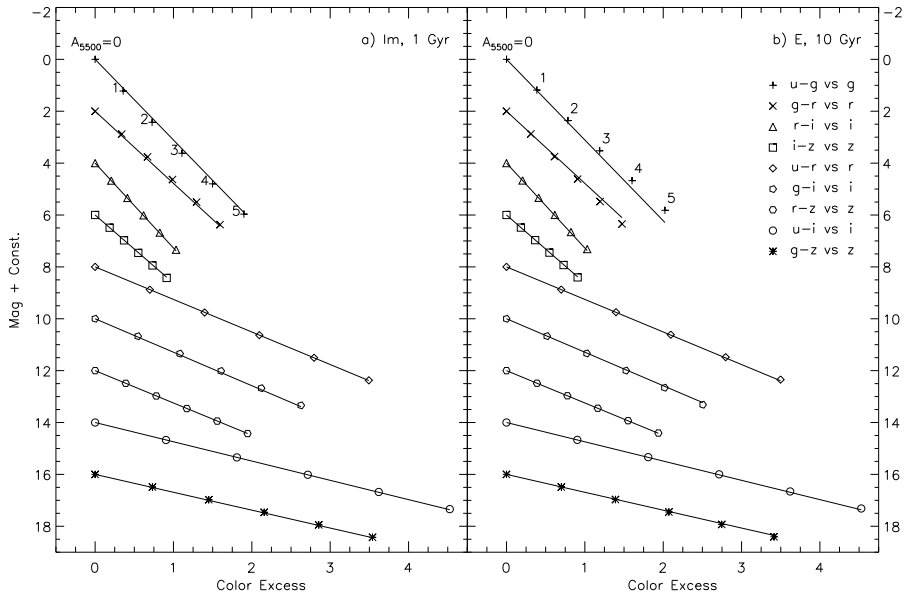}
\caption
{\label{fig:vectorg}The symbols show the actual reddening curves
of 1~Gyr ($a$) and 10~Gyr ($b$) galaxies at redshift $z=0$ for
nine different color-magnitude diagrams of SDSS filters.  The distance
between adjacent symbols in each curve corresponds to the extinction of one
magnitude at 5500~{\AA} ($A_{5500}=1$).  The lines show the ``reference''
reddening vectors of the corresponding filter set, whose slope is obtained
with the reference extinction law in Table \ref{table:Ai}.
The largest deviations from the linear relation are for the $u-g$ and $g-r$
colors, and are in a different sense for 1 Gyr vs. 10 Gyr galaxy SEDs.
The largest deviations from the reference reddening vector are also for
the $u-g$ and $g-r$ colors and for the elliptical galaxy.}
\end{figure*}

Figure \ref{fig:vectorg} shows the SED dependence and nonlinearity of the
extinction in two sample CM diagrams of nine different pairs of SDSS filters.
The discrepancies between the actual (symbol) and reference
(line) reddenings are generally larger 1) for the filter pairs involving
the $g$ filter, and 2) for the elliptical galaxy.  The former is
because the $g$ filter has a much larger $\Delta \lambda / \lambda_{c}$ ratio
than the other filters (see Table \ref{table:lambda}), and this causes the
central wavelength of $g$ to shift more significantly than the others as the
extinction becomes larger.  The fact that the reddening becomes
noticeably nonlinear when one of the two filters has a large $\Delta \lambda /
\lambda_{c}$ was also observed in the near-infrared CM diagrams (see Kim et al.
2005, 2006).

\begin{figure*}
\epsscale{0.8}
\plotone{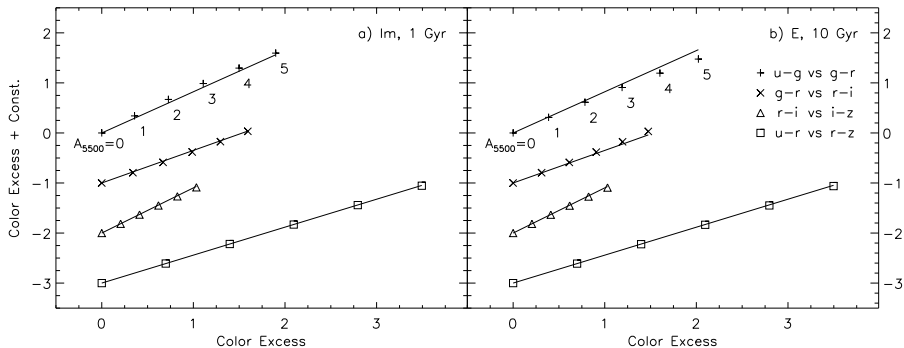}
\caption
{\label{fig:vectorccg}Same as Figure \ref{fig:vectorg}, but for four different
color-color diagrams.}
\end{figure*}

The reason that the elliptical galaxy shows larger discrepancies
is a bit more complex.  In the wavelength regime of the $g$ filter,
the elliptical galaxy has nearly flat or increasing $f_\lambda$ with
increasing $\lambda$ while the irregular galaxy has decreasing $f_\lambda$
(see Fig. \ref{fig:galspec}).
Thus, in the limit of small extinction, the effective wavelengths,
\begin{equation}
\label{lambda_eff}
	\lambda_{eff} = \exp \frac{\int d(\ln \nu) \, f_\nu S_\nu \ln \lambda}
	                         {\int d(\ln \nu) \, f_\nu S_\nu},
\end{equation}
of the elliptical galaxy are longer than $\lambda_c$, and, for example,
the $u-g$ ($g-r$) reddening curve of the elliptical galaxy is
shallower (steeper) than the reference reddening vector.  Conversely,
for the irregular galaxy, the $u-g$ ($g-r$) reddening curve is steeper
(shallower) than the reference reddening vector.  As the extinction becomes
more significant, $\lambda_{eff}$ values shift to longer wavelengths,
and the reddening curves for $u-g$ ($g-r$) become shallower (steeper)
for both elliptical and irregular galaxies.  Therefore, as the extinction
increases, the discrepancies between the actual and reference reddenings
for both $u-g$ and $g-r$ colors become larger for the elliptical galaxy
but smaller for the irregular galaxy.

Figure \ref{fig:vectorccg} shows the SED-dependence and nonlinearity of
the extinction in CC diagrams for four different color combinations of SDSS
filters.  As in Figure \ref{fig:vectorg}, the discrepancies between the
actual and reference reddenings are more significant for the color
combinations involving the $g$ filter and for the elliptical galaxy.
The behavior of the actual reddening curve relative to the reference
reddening vector can be explained similarly to that in the CM diagrams.
As an example, here we describe the reddening behavior in the
$u-g$ vs. $g-r$ diagram in detail.  In the limit of small extinction for
the elliptical galaxy, the actual $u-g$ ($g-r$) color
excess is larger (smaller) than the reference color excess (because
$\lambda_{eff}-\lambda_c$ is larger for $g$), and this makes the
actual reddening curve in $u-g$ vs. $g-r$ CC diagram shallower than
the reference reddening vector.  As the extinction becomes larger, the
$\lambda_{eff}$ of $g$ increases more rapidly than that of $u$ and $r$,
and the $u-g$ ($g-r$) color excess becomes larger (smaller) more quickly
compared to the reference color excess.  This makes the actual
reddening curve in the $u-g$ vs. $g-r$ diagram deviate further
from the reference vector as the extinction increases.

\begin{figure*}
\epsscale{1.0}
\plotone{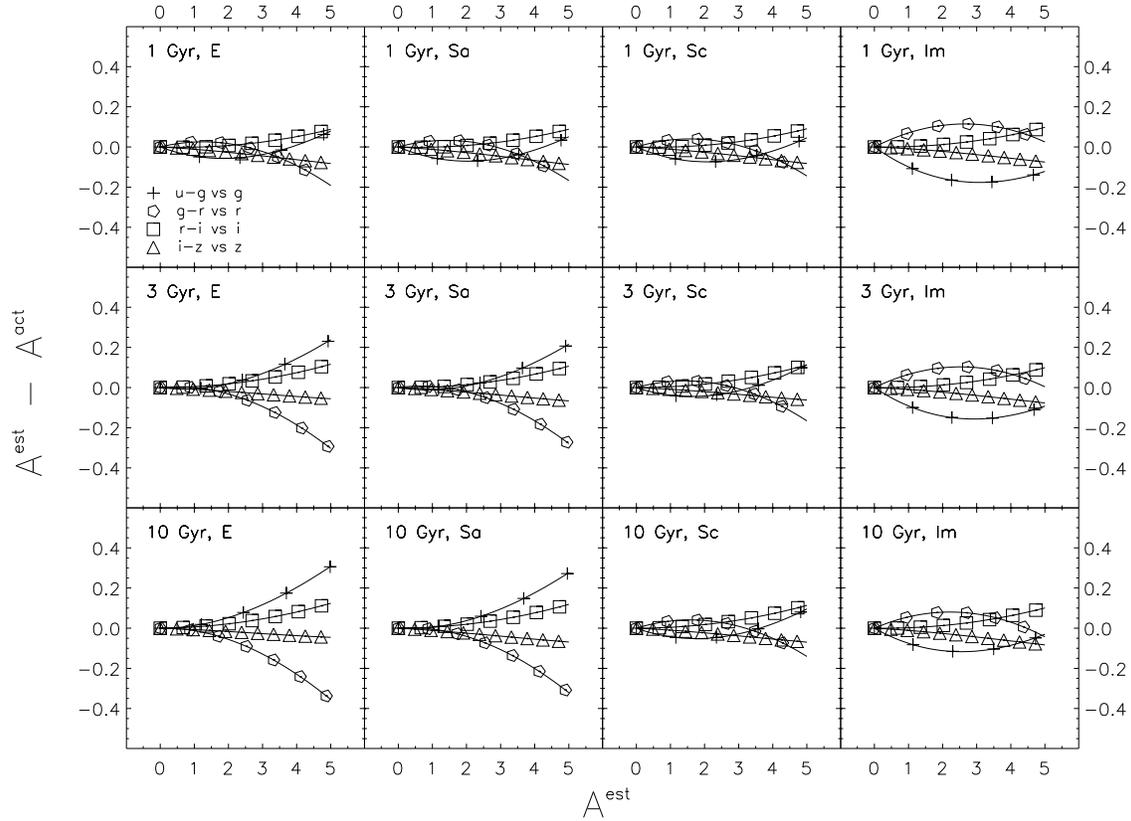}
\caption
{\label{fig:adiffag}The differences between the estimated extinction $A^{est}$
and the actual extinction $A^{act}$ in four different color-magnitude
diagrams for our $z=0$ galaxy models (four morphologies and three ages).
See the text for the definitions of $A^{est}$ and $A^{act}$.  The distance
between adjacent symbols in each curve corresponds to the extinction of one
magnitude at 5500~{\AA}.}
\end{figure*}

\begin{figure*}
\epsscale{1.0}
\plotone{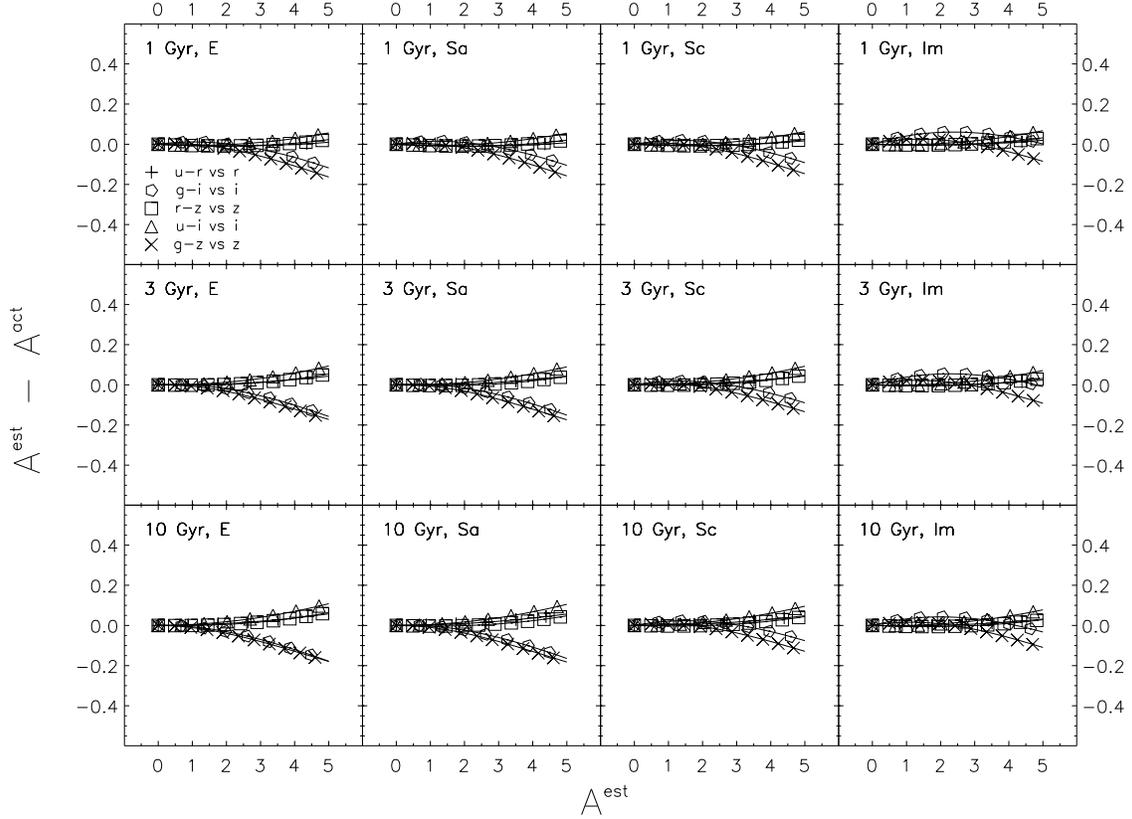}
\caption
{\label{fig:adiffbg}Same as Figure \ref{fig:adiffag}, but for five more
color-magnitude diagrams.}
\end{figure*}

We now discuss the way the SED dependence and nonlinearity of the
extinction cause errors in estimating the amount of extinction from
the observed colors.  Figures \ref{fig:adiffag}--\ref{fig:adiffbg} show
the differences between $A^{est}$ and $A^{act}$ in nine colors for our
1, 3, and 10~Gyr galaxy models.  $A^{est}$ is the amount of
extinction estimated from the calculated color excess and our reference
extinction law in Table~\ref{table:Ai}.  That is, $A^{est}$ is estimated by
\begin{equation}
\label{A_est}
        A_Y^{est} = \frac{(m_X-m_Y)-(m_X-m_Y)_0}{A_X^{ref}/A_Y^{ref}-1},
\end{equation}
where $m_X$ and $m_Y$ are the reddened magnitudes of two filters, $X$ and $Y$,
the subscript 0 denotes the intrinsic value, and $A_X^{ref}/A_Y^{ref}$ is
our reference extinction law between the two filters.
The figures show that in general, the
discrepancies between $A^{est}$ and $A^{act}$ are larger for older isochrones
and colors involving the $g$ filter, as expected from Figure \ref{fig:vectorg}.
This implies, for example, that applying to late-type
galaxies a reddening curve calculated for an early-type galaxy may
cause a non-negligible error in dereddening, particularly for colors involving
$g$.  Moreover, although the overall $A^{est}-A^{act}$ values are relatively
small for colors involving the $g$ filter at 1~Gyr, their
reddening curves are nonlinear at all ages.  
The largest relative difference of $|A^{est}-A^{act}|$ between different
10~Gyr galaxies is 9~\% and occurs in $u-g$
between the elliptical and irregular galaxies at $A^{est}_{g} \sim 1$.

\begin{figure*}
\epsscale{1.0}
\plotone{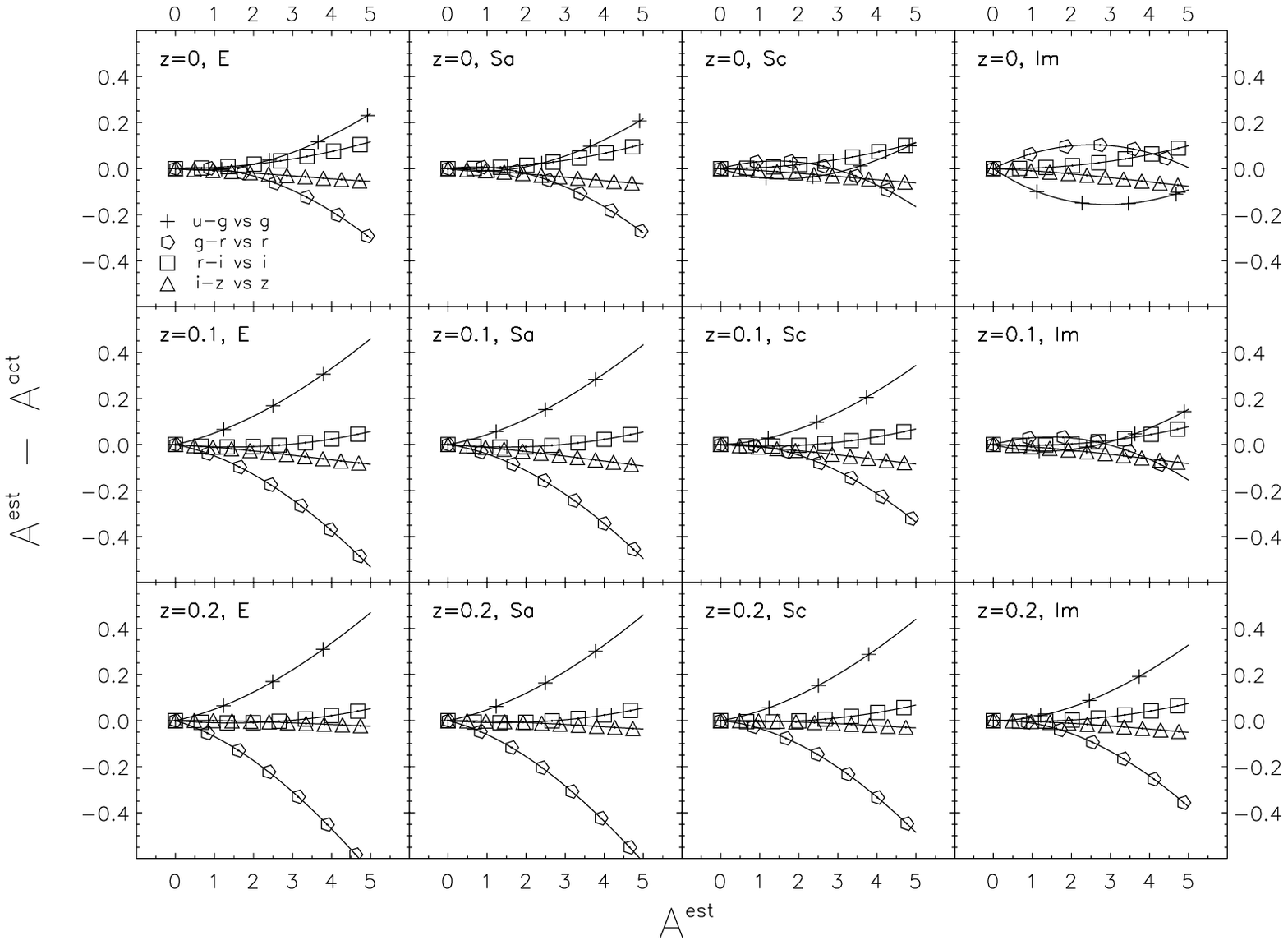}
\caption
{\label{fig:adiffagz}Same as Figure \ref{fig:adiffag}, but for our 3~Gyr
galaxy models with redshift values of 0, 0.1, and 0.2 in four different
SDSS colors.}
\end{figure*}

It is worth noting that the colors between $r$, $i$ and $z$
show very little SED dependence and nonlinearity.  This is because
these filters are located near the Rayleigh-Jeans regime of the stellar
spectra, and thus the colors between them are almost identical for all SEDs.

The colors involving two filters whose $\lambda_c$'s are further apart
exhibit smaller SED dependence and nonlinearity (e.g., $u-i$ has smaller
SED dependence and nonlinearity of extinction than $u-g$ and $u-r$).
This is because the shift of $\lambda_{eff}$ due to extinction becomes
relatively less significant as the gap between the two $\lambda_{eff}$
(or $\lambda_c$) values increases.

\begin{figure}
\epsscale{0.9}
\plotone{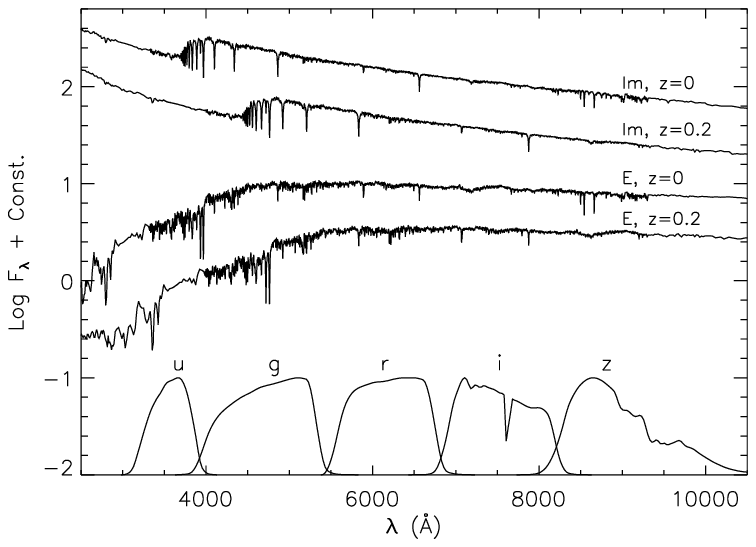}
\caption
{\label{fig:galspecz}Spectra of elliptical and irregular galaxies at
3~Gyr for two redshift values $z=0$ and 0.2.  Also shown are the
response functions of the SDSS filters for 1.3 airmasses.}
\end{figure}

Figure \ref{fig:adiffagz} shows the $A^{est}-A^{act}$
values for three different redshift values ($z=0$, 0.1, and 0.2) of four
galaxy morphologies at 3~Gyr.  The dependence of reddening curves on $z$
is not negligible for $z$ values between 0 and 0.2.  Again, the colors
involving the $g$ filter show the largest variations of $A^{est}-A^{act}$
between different $z$ values.  This is because the dependence of the SED
slope on $z$ is the largest in the $g$ filter, which is due to the Balmer
break (see Fig. \ref{fig:galspecz}).  As $z$ increases, the SED slope in the
$g$ filter increases to a more positive value and $\lambda_{eff}$
shifts to a longer wavelength.  Thus in case of $u-g$ vs.
$g$ ($g-r$ vs. $r$) CM diagram, the reddening vector becomes shallower
(steeper) and $A^{est}-A^{act}$ becomes larger (smaller).

The procedure for calculating the correct amount of extinction from an
observed color excess is then 1) to obtain $A^{est}$ from an observed color
excess and our reference extinction law in Table~\ref{table:Ai}, and 2) to
correct it with the corresponding $A^{est}-A^{act}$ value in Figures
\ref{fig:adiffag} to \ref{fig:adiffagz}, i.e. to subtract $A^{est}-A^{act}$
from $A^{est}$.\footnote{An electronic table of $A^{est}-A^{act}$ values is
available from the authors upon request.}
As our $A^{est}$ and $A^{act}$ values are calculated with the redshifted
spectra, any $K$-corrections can be applied independently from our
dereddening procedures.

\begin{figure*}
\epsscale{0.9}
\plotone{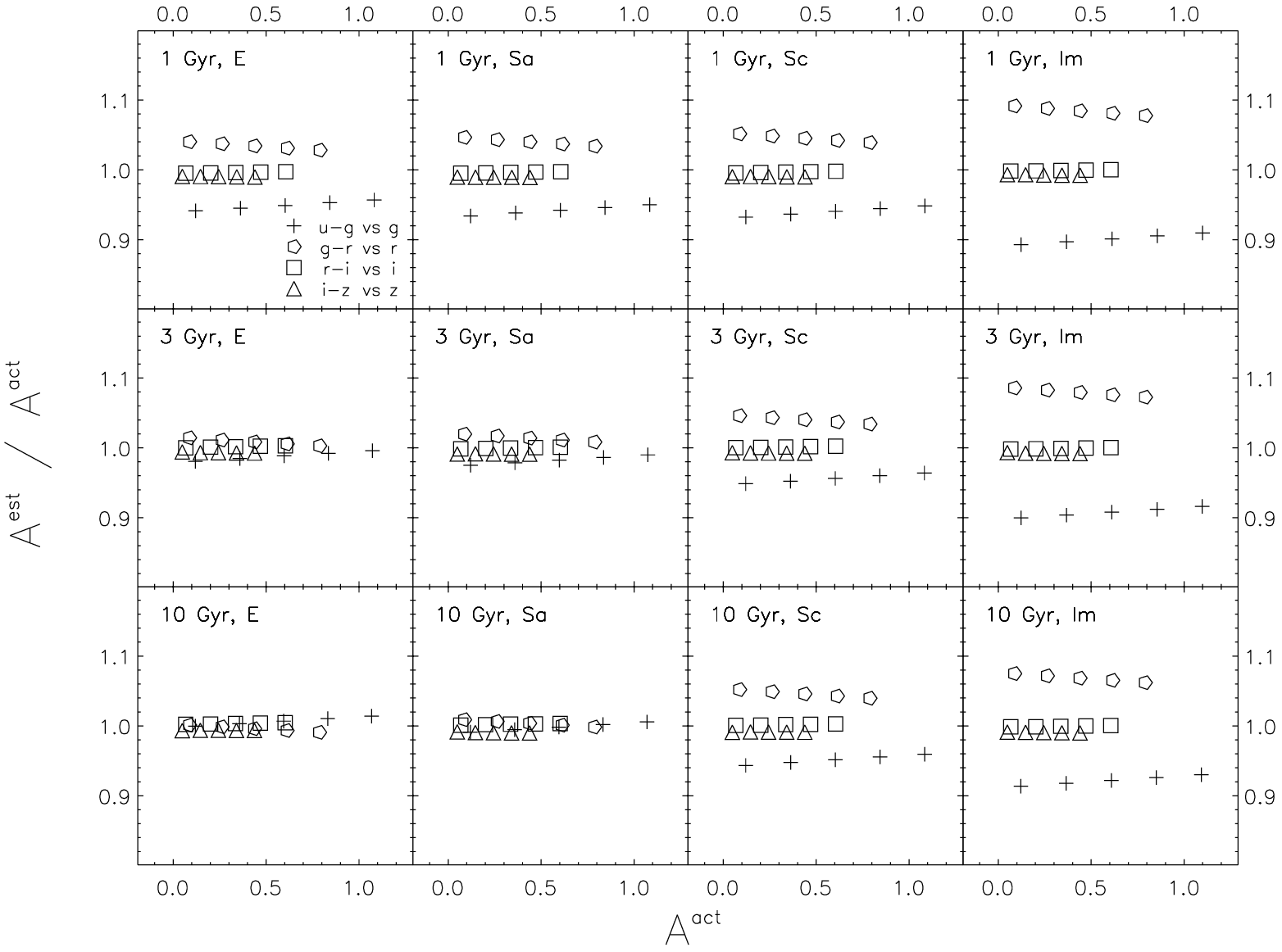}
\caption
{\label{fig:adiffagr}The ratios between the estimated extinction $A^{est}$
and the actual extinction $A^{act}$ in four different color-magnitude
diagrams for our $z=0$ galaxy models (four morphologies and three ages).
See the text for the definitions of $A^{est}$ and $A^{act}$.  This plot
is to show the SED-dependence of reddening at low extinction.  Each symbol,
from left ro right, represents extinction values of 0.1, 0.3, 0.5, 0.7, and
0.9 mag at 5500~{\AA}.}
\end{figure*}

Figures \ref{fig:adiffag} through \ref{fig:adiffagz} cannot effectively
show the SED-dependence of the reddening at low extinction.  This can be
manifested with $A^{est}/A^{act}$, instead of $A^{est}-A^{act}$.  Figure
\ref{fig:adiffagr} clearly shows that the SED-dependence of the reddening
is present even at low extinction and that the $A^{est}/A^{act}$ ratio is
a very weak function of extinction.

\begin{figure*}
\epsscale{1.0}
\plotone{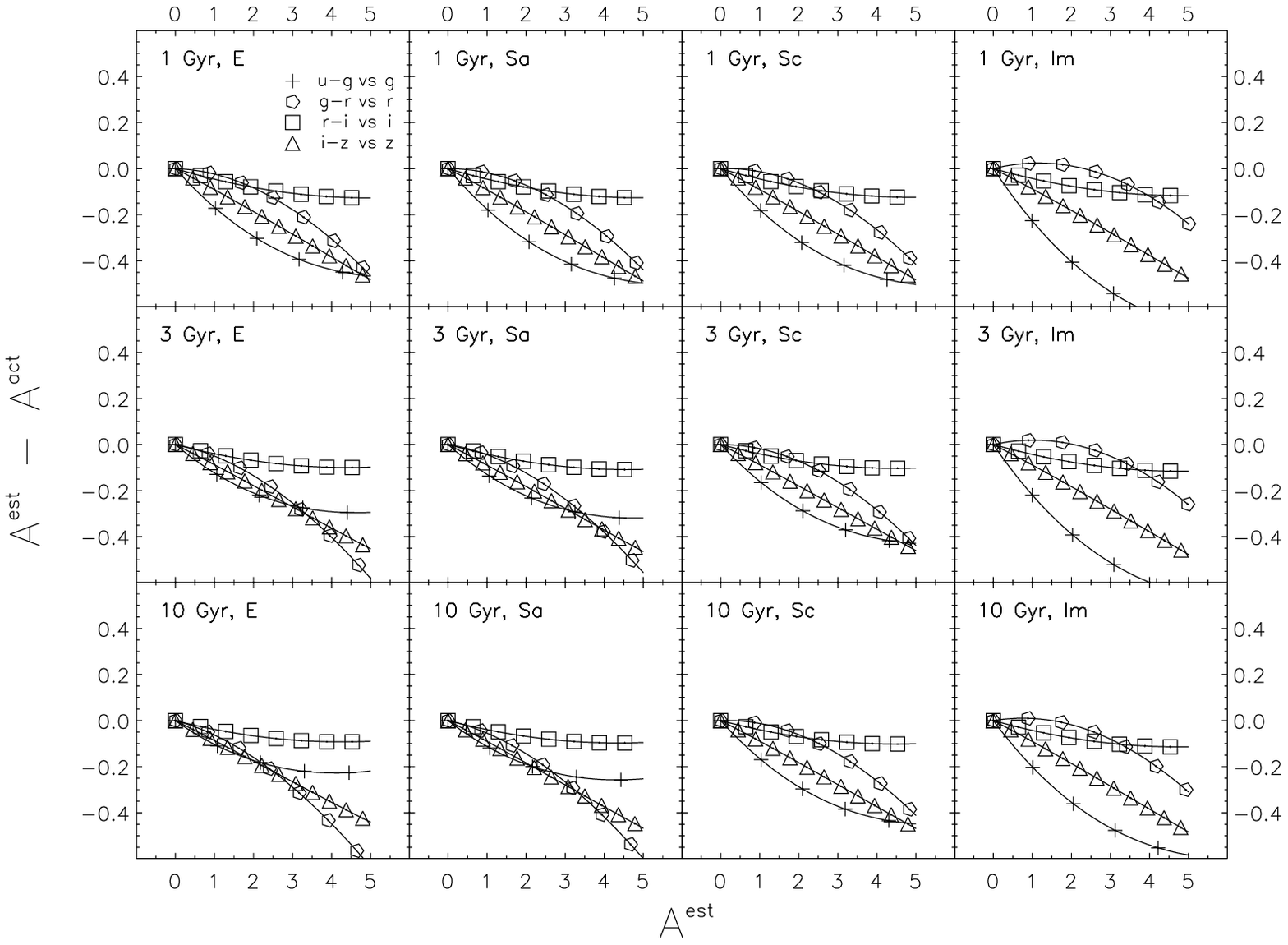}
\caption
{\label{fig:adiffsg}Same as Figure \ref{fig:adiffag}, but $A^{est}$ values
are calculated with the relative extinctions by Schlegel et al. (1998)
instead of our reference extinction law.}
\end{figure*}

Schlegel et al. (1998) gives the relative extinctions for the SDSS filters
that are calculated for a $z=0$ elliptical galaxy SED using the SDSS filter
response functions published by Fukugita et al. (1996).
These response functions are somewhat different from the most recent ones
available, which are the ones adopted in the present work, and result
in relative extinction values that are different from our reference
extinction law by upto $\sim 7$~\% (Table \ref{table:Ai}; the difference
is largest in $z$ while smaller than 3~\% in $u$ and $g$).  Figure
\ref{fig:adiffsg} shows the $A^{est}-A^{act}$ values where $A^{est}$ are
calculated with the relative extinctions by Schlegel et al. instead of
our reference extinction law.  Most of the $A^{est}-A^{act}$ values are
negative, meaning that using the relative extinctions by Schlegel et al.
would underestimate the amount of extinction in most cases by $\sim 5$ to
10~\% (maximum $\sim 20$~\%).  This is because the relative extinctions
by Schlegel et al. have a stronger wavelength dependence compared to our
values (see eq. \ref{A_est}).

\subsection{Extinction Values in the SDSS Archive}

The SDSS archive provides the amount of Galactic extinction in $r$ magnitude
at the position of each object that is computed following the relative
extinction by Schlegel et al. (1998; Table \ref{table:Ai}).  For the reasons
discussed above, the relative extinction depends on the SED of an object and
on the amount of extinction, which is not considered in the extinction values
given in the SDSS archive.  In Tables \ref{table:afit1} to \ref{table:afit3},
we provide the coefficients of the best-fit polynomials to $A_i$ as a function
of $A_{5500}$ for our 3 and 10 Gyr galaxy models (valid for $A_{5500}
\le 5$~mag).  The fitting polynomial has a form of
\begin{equation}
\label{afit}
	A_i = c_1 A_{5500} + c_2 (A_{5500})^2 + c_3 (A_{5500})^3.
\end{equation}
In order to correct the extinction value given in the SDSS archive, one first
divides the $r$ extinction value of the archive by 0.843 to obtain
$A_{5500}$ (that is, $A_V$), and use equation (\ref{afit}) to obtain $A_i$.
Here, 0.843 is the $A_{r}/A_V$ value
calculated by Schlegel et al. (1998) using the SDSS filter response
functions by Fukugita et al. (1996).
We find that the largest correction in $r$ for $A_{5500}=2.5$~mag made
by this procedure is 0.09~mag, and most of the correction comes
from the new filter response functions that we use.  The SED-dependence and
nonlinearity of the extinction results in corrections in $r$ of less than
0.035~mag at $A_{5500}=2.5$~mag, which is much less than the corrections
for the derredening from the color excesses discussed above.
This shows that small differences
in the relative extinction (i.e., extinction law) result in relatively
large differences in the estimated extinction from the color excess.

\begin{deluxetable*}{cccrrrcrrr}
\tabletypesize{\scriptsize}
\tablecolumns{10}
\tablewidth{0pt}
\tablecaption{
\label{table:afit1}Coefficients of Best-Fit Polynomials to $A_i$ as a Function
of $A_{5500}$ for $z$=0 Galaxies\tablenotemark{a}}
\tablehead{
\colhead{} &
\colhead{} &
\colhead{} &
\multicolumn{3}{c}{3 Gyr} &
\colhead{} &
\multicolumn{3}{c}{10 Gyr} \\ \cline{4-6} \cline{8-10}
\colhead{Morphology} &
\colhead{Filter} &
\colhead{} &
\colhead{$c_1$} &
\colhead{$c_2$} &
\colhead{$c_3$} &
\colhead{} &
\colhead{$c_1$} &
\colhead{$c_2$} &
\colhead{$c_3$}
}
\startdata
 E & $u$ & \hspace{0.2cm} &    1.574 & $-$1.25E-03 &    1.33E-05 & \hspace{0.2cm} &    1.574 & $-$1.20E-03 &    1.17E-05 \\
 E & $g$ & \hspace{0.2cm} &    1.197 & $-$6.52E-03 &    1.46E-04 & \hspace{0.2cm} &    1.192 & $-$6.25E-03 &    1.41E-04 \\
 E & $r$ & \hspace{0.2cm} &    0.877 & $-$1.44E-03 &    6.25E-06 & \hspace{0.2cm} &    0.876 & $-$1.44E-03 &    7.21E-06 \\
 E & $i$ & \hspace{0.2cm} &    0.672 & $-$1.57E-03 &    1.40E-06 & \hspace{0.2cm} &    0.671 & $-$1.56E-03 &    1.35E-06 \\
 E & $z$ & \hspace{0.2cm} &    0.487 & $-$1.03E-03 & $-$3.36E-06 & \hspace{0.2cm} &    0.487 & $-$1.04E-03 & $-$3.19E-06 \\
Sa & $u$ & \hspace{0.2cm} &    1.574 & $-$1.27E-03 &    1.36E-05 & \hspace{0.2cm} &    1.575 & $-$1.23E-03 &    1.32E-05 \\
Sa & $g$ & \hspace{0.2cm} &    1.199 & $-$6.63E-03 &    1.48E-04 & \hspace{0.2cm} &    1.195 & $-$6.40E-03 &    1.44E-04 \\
Sa & $r$ & \hspace{0.2cm} &    0.877 & $-$1.44E-03 &    6.61E-06 & \hspace{0.2cm} &    0.876 & $-$1.44E-03 &    7.23E-06 \\
Sa & $i$ & \hspace{0.2cm} &    0.672 & $-$1.56E-03 &    6.60E-07 & \hspace{0.2cm} &    0.671 & $-$1.57E-03 &    2.02E-06 \\
Sa & $z$ & \hspace{0.2cm} &    0.488 & $-$1.03E-03 & $-$3.14E-06 & \hspace{0.2cm} &    0.487 & $-$1.03E-03 & $-$3.03E-06 \\
Sc & $u$ & \hspace{0.2cm} &    1.577 & $-$1.38E-03 &    1.57E-05 & \hspace{0.2cm} &    1.578 & $-$1.39E-03 &    1.53E-05 \\
Sc & $g$ & \hspace{0.2cm} &    1.209 & $-$7.07E-03 &    1.51E-04 & \hspace{0.2cm} &    1.211 & $-$7.14E-03 &    1.50E-04 \\
Sc & $r$ & \hspace{0.2cm} &    0.878 & $-$1.45E-03 &    6.87E-06 & \hspace{0.2cm} &    0.878 & $-$1.45E-03 &    6.94E-06 \\
Sc & $i$ & \hspace{0.2cm} &    0.673 & $-$1.57E-03 &    1.87E-06 & \hspace{0.2cm} &    0.673 & $-$1.57E-03 &    1.05E-06 \\
Sc & $z$ & \hspace{0.2cm} &    0.488 & $-$1.03E-03 & $-$2.48E-06 & \hspace{0.2cm} &    0.488 & $-$1.02E-03 & $-$2.84E-06 \\
Im & $u$ & \hspace{0.2cm} &    1.579 & $-$1.45E-03 &    1.52E-05 & \hspace{0.2cm} &    1.579 & $-$1.44E-03 &    1.61E-05 \\
Im & $g$ & \hspace{0.2cm} &    1.226 & $-$7.62E-03 &    1.44E-04 & \hspace{0.2cm} &    1.221 & $-$7.48E-03 &    1.47E-04 \\
Im & $r$ & \hspace{0.2cm} &    0.881 & $-$1.47E-03 &    5.77E-06 & \hspace{0.2cm} &    0.880 & $-$1.46E-03 &    6.15E-06 \\
Im & $i$ & \hspace{0.2cm} &    0.675 & $-$1.56E-03 & $-$1.06E-06 & \hspace{0.2cm} &    0.674 & $-$1.56E-03 & $-$4.35E-08 \\
Im & $z$ & \hspace{0.2cm} &    0.490 & $-$9.92E-04 & $-$4.02E-06 & \hspace{0.2cm} &    0.490 & $-$9.97E-04 & $-$3.87E-06 \\
\enddata
\tablenotetext{a}{The coefficients are to be used with the fitting formula of equation (\ref{afit}), which is valid for $A_{5500} \le 5$~mag.}
\end{deluxetable*}

\begin{deluxetable*}{cccrrrcrrr}
\tabletypesize{\scriptsize}
\tablecolumns{10}
\tablewidth{0pt}
\tablecaption{
\label{table:afit2}Coefficients of Best-Fit Polynomials to $A_i$ as a Function
of $A_{5500}$ for $z$=0.1 Galaxies\tablenotemark{a}}
\tablehead{
\colhead{} &
\colhead{} &
\colhead{} &
\multicolumn{3}{c}{3 Gyr} &
\colhead{} &
\multicolumn{3}{c}{10 Gyr} \\ \cline{4-6} \cline{8-10}
\colhead{Morphology} &
\colhead{Filter} &
\colhead{} &
\colhead{$c_1$} &
\colhead{$c_2$} &
\colhead{$c_3$} &
\colhead{} &
\colhead{$c_1$} &
\colhead{$c_2$} &
\colhead{$c_3$}
}
\startdata
 E & $u$ & \hspace{0.2cm} &    1.572 & $-$1.16E-03 &    1.22E-05 & \hspace{0.2cm} &    1.569 & $-$1.09E-03 &    1.13E-05 \\
 E & $g$ & \hspace{0.2cm} &    1.178 & $-$5.88E-03 &    1.49E-04 & \hspace{0.2cm} &    1.172 & $-$5.65E-03 &    1.51E-04 \\
 E & $r$ & \hspace{0.2cm} &    0.874 & $-$1.42E-03 &    8.13E-06 & \hspace{0.2cm} &    0.873 & $-$1.41E-03 &    8.58E-06 \\
 E & $i$ & \hspace{0.2cm} &    0.672 & $-$1.59E-03 &    7.55E-07 & \hspace{0.2cm} &    0.671 & $-$1.60E-03 &    1.29E-06 \\
 E & $z$ & \hspace{0.2cm} &    0.488 & $-$1.02E-03 & $-$2.81E-06 & \hspace{0.2cm} &    0.488 & $-$1.02E-03 & $-$3.23E-06 \\
Sa & $u$ & \hspace{0.2cm} &    1.573 & $-$1.19E-03 &    1.29E-05 & \hspace{0.2cm} &    1.571 & $-$1.15E-03 &    1.17E-05 \\
Sa & $g$ & \hspace{0.2cm} &    1.181 & $-$6.02E-03 &    1.52E-04 & \hspace{0.2cm} &    1.176 & $-$5.85E-03 &    1.53E-04 \\
Sa & $r$ & \hspace{0.2cm} &    0.875 & $-$1.43E-03 &    8.71E-06 & \hspace{0.2cm} &    0.874 & $-$1.41E-03 &    8.33E-06 \\
Sa & $i$ & \hspace{0.2cm} &    0.672 & $-$1.59E-03 &    1.35E-06 & \hspace{0.2cm} &    0.671 & $-$1.60E-03 &    1.68E-06 \\
Sa & $z$ & \hspace{0.2cm} &    0.488 & $-$1.01E-03 & $-$3.78E-06 & \hspace{0.2cm} &    0.488 & $-$1.01E-03 & $-$3.66E-06 \\
Sc & $u$ & \hspace{0.2cm} &    1.580 & $-$1.33E-03 &    1.40E-05 & \hspace{0.2cm} &    1.581 & $-$1.35E-03 &    1.38E-05 \\
Sc & $g$ & \hspace{0.2cm} &    1.194 & $-$6.45E-03 &    1.50E-04 & \hspace{0.2cm} &    1.197 & $-$6.54E-03 &    1.49E-04 \\
Sc & $r$ & \hspace{0.2cm} &    0.876 & $-$1.43E-03 &    7.11E-06 & \hspace{0.2cm} &    0.876 & $-$1.44E-03 &    7.32E-06 \\
Sc & $i$ & \hspace{0.2cm} &    0.673 & $-$1.58E-03 &    4.58E-07 & \hspace{0.2cm} &    0.673 & $-$1.58E-03 & $-$1.08E-07 \\
Sc & $z$ & \hspace{0.2cm} &    0.489 & $-$1.01E-03 & $-$3.26E-06 & \hspace{0.2cm} &    0.489 & $-$1.00E-03 & $-$4.42E-06 \\
Im & $u$ & \hspace{0.2cm} &    1.584 & $-$1.41E-03 &    1.52E-05 & \hspace{0.2cm} &    1.583 & $-$1.40E-03 &    1.56E-05 \\
Im & $g$ & \hspace{0.2cm} &    1.211 & $-$6.85E-03 &    1.36E-04 & \hspace{0.2cm} &    1.207 & $-$6.76E-03 &    1.41E-04 \\
Im & $r$ & \hspace{0.2cm} &    0.880 & $-$1.47E-03 &    6.83E-06 & \hspace{0.2cm} &    0.879 & $-$1.46E-03 &    7.61E-06 \\
Im & $i$ & \hspace{0.2cm} &    0.675 & $-$1.58E-03 & $-$1.49E-07 & \hspace{0.2cm} &    0.675 & $-$1.58E-03 &    3.68E-07 \\
Im & $z$ & \hspace{0.2cm} &    0.490 & $-$9.98E-04 & $-$3.60E-06 & \hspace{0.2cm} &    0.490 & $-$1.00E-03 & $-$3.31E-06 \\
\enddata
\tablenotetext{a}{The coefficients are to be used with the fitting formula of equation (\ref{afit}), which is valid for $A_{5500} \le 5$~mag.}
\end{deluxetable*}

\begin{deluxetable*}{cccrrrcrrr}
\tabletypesize{\scriptsize}
\tablecolumns{10}
\tablewidth{0pt}
\tablecaption{
\label{table:afit3}Coefficients of Best-Fit Polynomials to $A_i$ as a Function
of $A_{5500}$ for $z$=0.2 Galaxies\tablenotemark{a}}
\tablehead{
\colhead{} &
\colhead{} &
\colhead{} &
\multicolumn{3}{c}{3 Gyr} &
\colhead{} &
\multicolumn{3}{c}{10 Gyr} \\ \cline{4-6} \cline{8-10}
\colhead{Morphology} &
\colhead{Filter} &
\colhead{} &
\colhead{$c_1$} &
\colhead{$c_2$} &
\colhead{$c_3$} &
\colhead{} &
\colhead{$c_1$} &
\colhead{$c_2$} &
\colhead{$c_3$}
}
\startdata
 E & $u$ & \hspace{0.2cm} &    1.565 & $-$1.06E-03 &    1.28E-05 & \hspace{0.2cm} &    1.562 & $-$1.04E-03 &    1.27E-05 \\
 E & $g$ & \hspace{0.2cm} &    1.174 & $-$6.17E-03 &    1.75E-04 & \hspace{0.2cm} &    1.170 & $-$6.19E-03 &    1.80E-04 \\
 E & $r$ & \hspace{0.2cm} &    0.875 & $-$1.46E-03 &    7.98E-06 & \hspace{0.2cm} &    0.873 & $-$1.46E-03 &    9.01E-06 \\
 E & $i$ & \hspace{0.2cm} &    0.672 & $-$1.57E-03 &    4.38E-07 & \hspace{0.2cm} &    0.671 & $-$1.58E-03 &    1.48E-06 \\
 E & $z$ & \hspace{0.2cm} &    0.487 & $-$1.01E-03 & $-$3.28E-06 & \hspace{0.2cm} &    0.486 & $-$1.02E-03 & $-$2.27E-06 \\
Sa & $u$ & \hspace{0.2cm} &    1.568 & $-$1.13E-03 &    1.41E-05 & \hspace{0.2cm} &    1.567 & $-$1.13E-03 &    1.39E-05 \\
Sa & $g$ & \hspace{0.2cm} &    1.177 & $-$6.28E-03 &    1.75E-04 & \hspace{0.2cm} &    1.174 & $-$6.32E-03 &    1.80E-04 \\
Sa & $r$ & \hspace{0.2cm} &    0.875 & $-$1.46E-03 &    8.09E-06 & \hspace{0.2cm} &    0.874 & $-$1.47E-03 &    9.69E-06 \\
Sa & $i$ & \hspace{0.2cm} &    0.672 & $-$1.57E-03 &    8.19E-07 & \hspace{0.2cm} &    0.672 & $-$1.57E-03 &    5.86E-07 \\
Sa & $z$ & \hspace{0.2cm} &    0.487 & $-$1.01E-03 & $-$3.18E-06 & \hspace{0.2cm} &    0.486 & $-$1.01E-03 & $-$2.68E-06 \\
Sc & $u$ & \hspace{0.2cm} &    1.579 & $-$1.35E-03 &    1.50E-05 & \hspace{0.2cm} &    1.581 & $-$1.37E-03 &    1.47E-05 \\
Sc & $g$ & \hspace{0.2cm} &    1.187 & $-$6.55E-03 &    1.73E-04 & \hspace{0.2cm} &    1.190 & $-$6.66E-03 &    1.74E-04 \\
Sc & $r$ & \hspace{0.2cm} &    0.877 & $-$1.47E-03 &    7.63E-06 & \hspace{0.2cm} &    0.877 & $-$1.48E-03 &    7.42E-06 \\
Sc & $i$ & \hspace{0.2cm} &    0.673 & $-$1.58E-03 &    5.63E-07 & \hspace{0.2cm} &    0.673 & $-$1.57E-03 &    3.32E-07 \\
Sc & $z$ & \hspace{0.2cm} &    0.488 & $-$1.00E-03 & $-$3.81E-06 & \hspace{0.2cm} &    0.488 & $-$1.01E-03 & $-$3.19E-06 \\
Im & $u$ & \hspace{0.2cm} &    1.584 & $-$1.41E-03 &    1.56E-05 & \hspace{0.2cm} &    1.583 & $-$1.39E-03 &    1.46E-05 \\
Im & $g$ & \hspace{0.2cm} &    1.199 & $-$6.82E-03 &    1.68E-04 & \hspace{0.2cm} &    1.196 & $-$6.75E-03 &    1.70E-04 \\
Im & $r$ & \hspace{0.2cm} &    0.881 & $-$1.48E-03 &    5.95E-06 & \hspace{0.2cm} &    0.880 & $-$1.48E-03 &    6.59E-06 \\
Im & $i$ & \hspace{0.2cm} &    0.676 & $-$1.57E-03 & $-$4.14E-07 & \hspace{0.2cm} &    0.675 & $-$1.57E-03 & $-$3.46E-07 \\
Im & $z$ & \hspace{0.2cm} &    0.490 & $-$9.94E-04 & $-$3.53E-06 & \hspace{0.2cm} &    0.489 & $-$9.95E-04 & $-$3.91E-06 \\
\enddata
\tablenotetext{a}{The coefficients are to be used with the fitting formula of equation (\ref{afit}), which is valid for $A_{5500} \le 5$~mag.}
\end{deluxetable*}

\begin{deluxetable*}{cccrrrrcrrrr}
\tabletypesize{\scriptsize}
\tablecolumns{12}
\tablewidth{0pt}
\tablecaption{
\label{table:ci}Intrinsic Colors of Our Galaxy Models\tablenotemark{a}}
\tablehead{
\colhead{} &
\colhead{} &
\colhead{} &
\multicolumn{4}{c}{3 Gyr} &
\colhead{} &
\multicolumn{4}{c}{10 Gyr} \\ \cline{4-7} \cline{9-12}
\colhead{Redshift} &
\colhead{Morphology} &
\colhead{} &
\colhead{$u-g$} &
\colhead{$g-r$} &
\colhead{$r-i$} &
\colhead{$i-z$} &
\colhead{} &
\colhead{$u-g$} &
\colhead{$g-r$} &
\colhead{$r-i$} &
\colhead{$i-z$}
}
\startdata
0.0 & E  & \hspace{0.2cm} &    1.51 &    0.69 &    0.32 &    0.24 & \hspace{0.2cm} &    1.66 &    0.80 &    0.38 &    0.28 \\
0.0 & Sa & \hspace{0.2cm} &    1.41 &    0.65 &    0.29 &    0.21 & \hspace{0.2cm} &    1.53 &    0.74 &    0.36 &    0.25 \\
0.0 & Sc & \hspace{0.2cm} &    1.07 &    0.48 &    0.24 &    0.19 & \hspace{0.2cm} &    1.00 &    0.45 &    0.24 &    0.18 \\
0.0 & Im & \hspace{0.2cm} &    0.64 &    0.11 &    0.07 &    0.06 & \hspace{0.2cm} &    0.76 &    0.23 &    0.13 &    0.09 \\
0.1 & E  & \hspace{0.2cm} &    1.56 &    0.87 &    0.35 &    0.28 & \hspace{0.2cm} &    1.70 &    1.00 &    0.41 &    0.34 \\
0.1 & Sa & \hspace{0.2cm} &    1.46 &    0.82 &    0.33 &    0.26 & \hspace{0.2cm} &    1.56 &    0.93 &    0.39 &    0.31 \\
0.1 & Sc & \hspace{0.2cm} &    1.05 &    0.59 &    0.26 &    0.22 & \hspace{0.2cm} &    0.97 &    0.55 &    0.25 &    0.22 \\
0.1 & Im & \hspace{0.2cm} &    0.64 &    0.18 &    0.06 &    0.07 & \hspace{0.2cm} &    0.75 &    0.30 &    0.13 &    0.12 \\
0.2 & E  & \hspace{0.2cm} &    1.61 &    1.14 &    0.43 &    0.29 & \hspace{0.2cm} &    1.80 &    1.30 &    0.49 &    0.34 \\
0.2 & Sa & \hspace{0.2cm} &    1.48 &    1.07 &    0.40 &    0.27 & \hspace{0.2cm} &    1.59 &    1.20 &    0.46 &    0.32 \\
0.2 & Sc & \hspace{0.2cm} &    0.89 &    0.78 &    0.31 &    0.22 & \hspace{0.2cm} &    0.80 &    0.72 &    0.29 &    0.22 \\
0.2 & Im & \hspace{0.2cm} &    0.48 &    0.33 &    0.08 &    0.06 & \hspace{0.2cm} &    0.59 &    0.46 &    0.15 &    0.12 \\
\enddata
\end{deluxetable*}

\subsection{Estimation of Morphology and Redshift from CC Diagrams}

\begin{figure*}
\epsscale{0.9}
\plotone{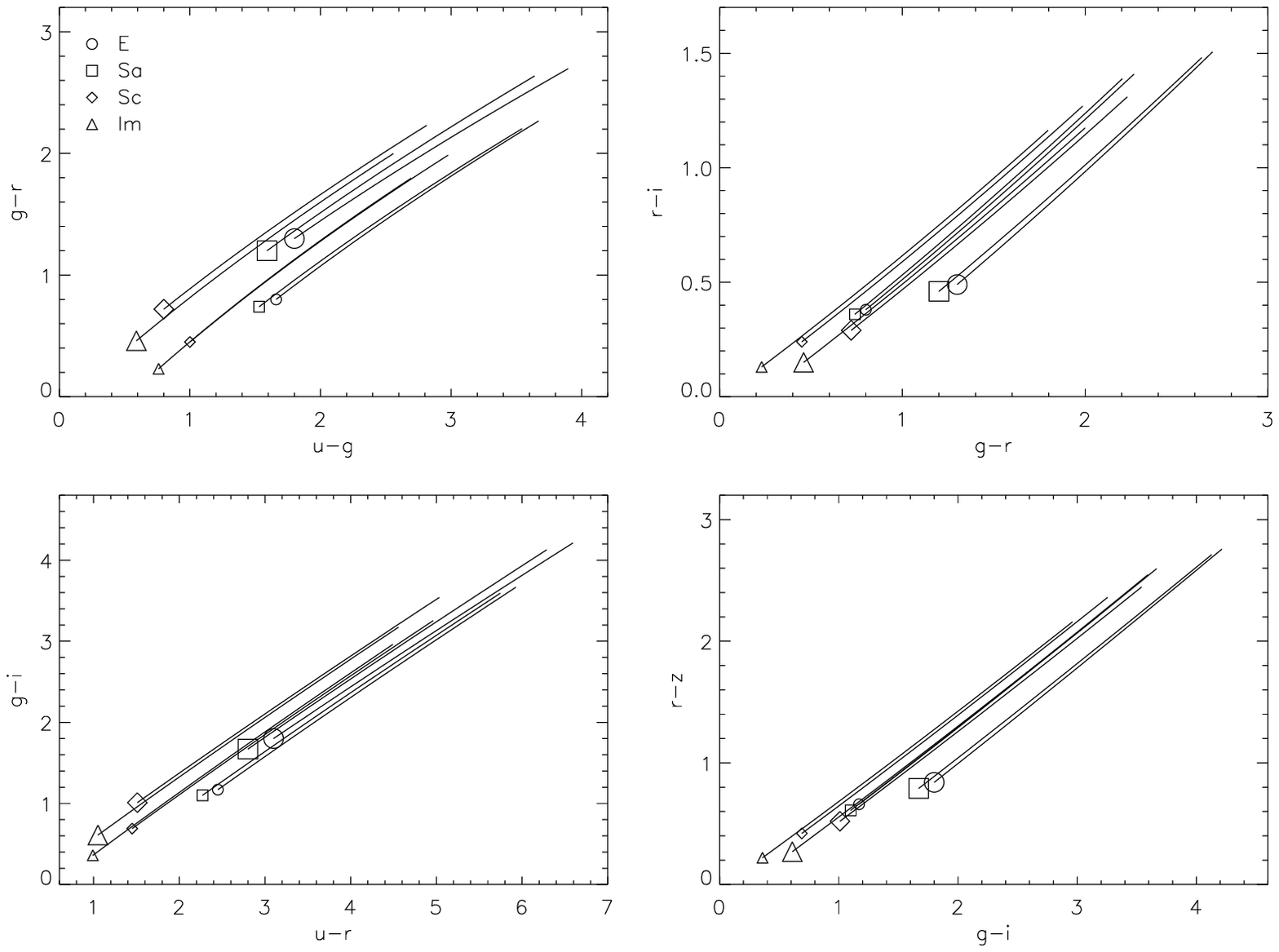}
\caption
{\label{fig:ccd}Intrinsic colors of our 10~Gyr galaxy models (symbols)
in four color-color diagrams along with their reddening curves (lines).
Small symbols are for $z=0$ models and large symbols for $z=0.2$.
The length of the reddening curve corresponds to $A_{5500}=5$~mag.}
\end{figure*}

As the reddening behavior is dependent on the morphology and redshift
of the galaxy, one needs to know this information before applying
our fitting formulae.  When visual inspection is not pratical or
feasible, one could use the Petrosian inverse concentration index
(Strateva et al. 2001, Shimasaku et al. 2001), S\'ersic index
(Blanton et al. 2003), Gini coefficient (Abraham, van den Bergh,
\& Nair 2003), coarseness parameter (Yamauchi et al. 2005),
or radial color gradient (Park \& Choi 2005) as a measure of the
morphology of galaxies.
Alternatively, one could estimate the morphology from the location
in the CC diagram.  Figure \ref{fig:ccd} shows the intrinsic colors of our
model galaxies along with the reddening curves.  There are cases where
one galaxy is located near the reddening curve of another galaxy of
different morphology.  In such cases, one would need an
approximate value of extinction based on the spatial location of the
galaxy in the sky from, e.g., Schlegel et al. (1998) to disentangle
the degeneracy.

\begin{figure*}
\epsscale{0.9}
\plotone{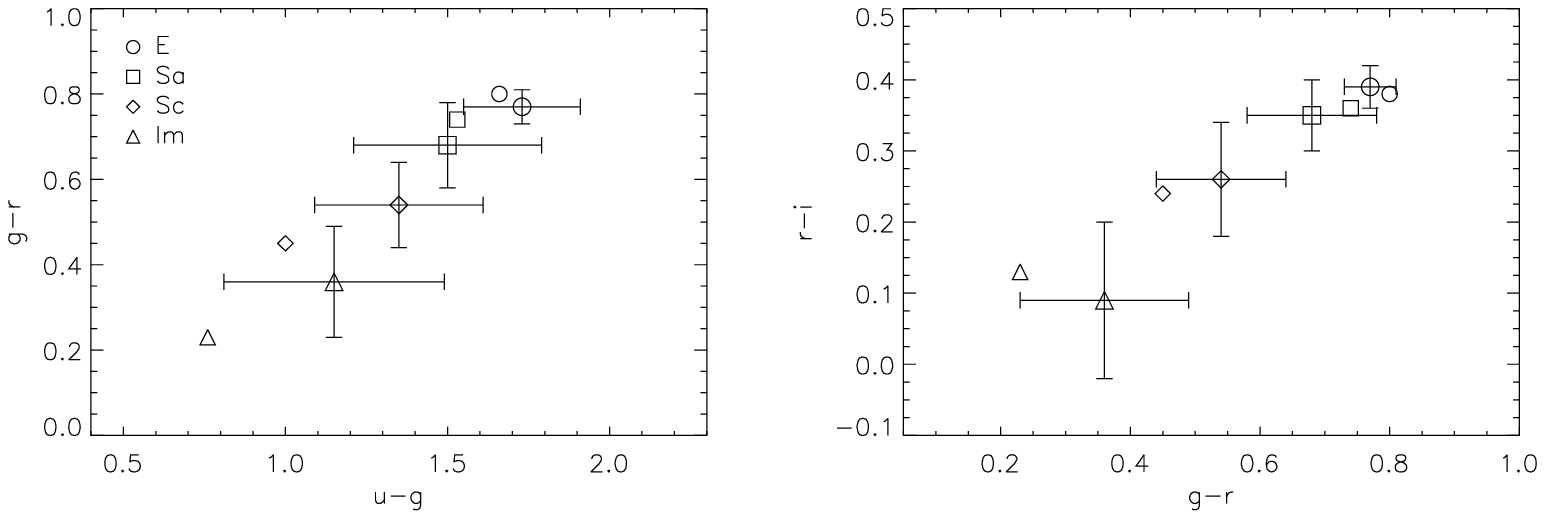}
\caption
{\label{fig:ccd2}Intrinsic colors of our 10~Gyr, $z=0$ galaxy models
(symbols) and the mean colors of visually classified SDSS galaxies after
$K$-correction by Fukugita et al. (2007; symbols with error bars).  Error
bars represent the dispersion.}
\end{figure*}

When the redshift of the galaxy is not known (e.g., when the spectrum is
not available), it needs to be estimated from the colors as well.
Figure \ref{fig:ccd} shows that the directions of the increasing reddening
and redshift are separated the most in $u-g$ vs. $g-r$ CC diagram for small
$z$ values ($\lsim 0.2$), primarily due to the Balmer break.
But in most cases, the redshift is degenerate with the reddening, and the
SED-dependence and nonlinearity of the reddening need to be incorporated
into the estimation of the redshift from photometric data.

Figure \ref{fig:ccd2} shows that our synthetic galaxy colors roughly agree
with the visually classified catalog of SDSS galaxies by Fukugita et al. (2007)
nearly within the dispersion of each morphology class.  Late-type galaxies
show larger discrepancies, probably because the population synthesis model
of Buzzoni (2005) does not include the effect of internal dust, which
suppresses ultraviolet emission.  This effect would result in bluer
colors in short-wavelength filters ($u-g$ \& $g-r$) for late-type
galaxies (Sc and Im), and this fact needs to be taken into account when
our fitting formulae are applied to visually classified galaxies.

\section{SUMMARY}
\label{sec:summary}

We have analyzed the behaviors of reddening curves in color-magnitude and
color-color diagrams of the SDSS photometric system for various galaxy models
with redshift values up to 0.2.

We have shown that the SED-dependence and the nonlinearity of the reddening
in the SDSS filters are not negligible.  These behaviors are the most
significant for colors involving the $g$ filter, which has the
largest $\Delta \lambda / \lambda_{c}$ value among SDSS filters.

The SDSS colors involving adjacent filters have greater SED-dependence
and nonlinearity because the shift of $\lambda_{eff}$ due to extinction
becomes relatively more significant as the gap between the two
$\lambda_{eff}$ (or $\lambda_c$) values decreases.

To provide a procedure to calculate the correct amount of extinction
from an observed color excess, we defined a ``reference extinction law''
that an imaginary filter having a Dirac delta transmission function at
$\lambda_c$ of an SDSS filter $i$ would experience for the SED of
a 10~Gyr elliptical galaxy when $A_{5500}=0.1$.
In order to calculate the amount of extinction for a given object,
one first obtains $A^{est}$ from an observed color excess and our
reference reddening vector, then corrects it with the corresponding
$A^{est}-A^{act}$ value in Figures \ref{fig:adiffag} to \ref{fig:adiffagz}.

The relative extinctions between (i.e., the extinction law for) SDSS filters
given by Schlegel et al., which were calculated with an older version of
filter response functions, would underestimate the amount of extinction
in most cases by $\sim 5$ to 10~\% (maximum $\sim 20$~\%).  Our reference
extinction law in Table \ref{table:Ai} is recommended instead when the
extinction is small (i.e., when the nonlinearity and SED-dependence are
not important, or $A < 1$).

Finally, we have shown that the dependence of reddening on redshift at
low extinction is the largest for colors involving the $g$ filter as well,
which is due to the Balmer break.

\acknowledgements
We thank Yun-Young Choi, Myungshin Im, Juhan Kim, Sang Chul Kim,
Hwankyung Sung, and Suk-Jin Yoon for helpful discussion.  We also deeply
thank the anonymous referee, whose comments greatly improved our manuscript.
S. S. K. was supported by the Astrophysical Research Center for
the Structure and Evolution of the Cosmos (ARCSEC) of the Korea Science and
Engineering Foundation through the Science Research Center (SRC) program.
M. G. L. was in part supported by ABRL (R14-2002-058-01000-0).
This work was in part supported by the BK21 program as well.


\end{document}